\documentstyle[aps,multicol,epsf]{revtex}

\begin{document}

\title{Sandpile Model for Relaxation in\\ Complex Systems}

\author{Alexei V\'azquez}

\address{Department of Theoretical Physics. Havana University. Havana
10400, Cuba.\\ICTP}

\author{Oscar Sotolongo-Costa}

\address{Department of Theoretical Physics. Havana University. Havana
10400, Cuba.}

\author{Francois Brouers}

\address{Department of Physics. University of Li\'ege. 4000. Belgium.}

\date{\today}

\maketitle

\begin{abstract}
The relaxation in complex systems is in general nonexponential. After an
initial rapid decay the system relaxes slowly following a long time tail. In
the present paper a sandpile modelation of the relaxation in complex systems
is analysed. Complexity is introduced by a process of avalanches in the
Bethe lattice and a feedback mechanism which leads to slower decay with
increasing time. In this way, some features of relaxation in complex
systems: long time tails relaxation, aging, and fractal distribution of
characteristic times, are obtained by simple computer simulations. 
\end{abstract}

\pacs{PACS numbers: 61.20.Lc, 64.60.Lx}

\begin{multicols}{2}

About 150 years ago Weber and Gauss carried out a simple experiment
demonstrating that relaxation in complex systems is not exponential.
Investigating the contraction of a silk thread they found that it does not
contract so quickly but it relaxes slowly following a power law $t^{-\alpha }$. This behavior is not particular of mechanical relaxation, it has been
also observed in experiments of magnetic relaxation in spin glasses and
high-Tc superconductors, transient current measurements in amorphous
semiconductors, dielectric relaxation, and more \cite{rel}. This phenomenon
of slow relaxation following a power law is called a {\em long time tail}
and it seems to be characteristic of the relaxation in complex systems.

Associated to the slow relaxation dynamics aging is observed,
which means that the properties of a system depends on its age \cite{bou}.
For instance, consider a glass quenched at time $t=0$ below its glass
transition temperature under an external stress and at time $t=t_w$ the
stress is released. If the system were near equilibrium then its response
measured at certain time $t>t_w$ will be a function of the difference $t-t_w$ and, therefore, normalized responses taken at different initial times $t_w$ will collapse in a single curve. However, in the glass material it is
observed that the relaxation becomes slower for larger $t_w$. The aging
phenomena is a consequence of the nonequilibrium dynamics and may be also
considered a characteristic of the dynamics of complex systems out from
equilibrium.

Sandpiles seem to be simplest systems which also lead to complex behavior..
They have been taken as the paradigm of self-organization since the
introduction of these ideas by Bak, Tang, and Wiesenfeld \cite{bak}. The
evolution of a sandpile under an external perturbation has been extensively
studied experimentally \cite{jae,hel}, theoretically \cite{tan,zap}, and by
computer simulations \cite{kad}. In general these works analyze the sandpile
dynamics close to the critical state, i.e. close to the critical angle.

Slow relaxation dynamics have been observed in the process of compactation
of sand heaps under vibrations \cite{kni,bar}. On the other hand, Jaeger 
{\it et al} have measured the temporal decay of the slope angle of sandpiles
under vibrations with different intensities \cite{jae}. They observed that
for high vibration intensities the pile angle relaxes logarithmically slowly to a steady state but it relaxes slower for low intensities. In order to
explain the slow dynamics they proposed a simple model, very similar to flux
creep models for magnetic relaxation in superconductors. In this way they
explain the logarithmic decay observed for high vibration intensities but
they could not account for the behavior at low intensities. The model
proposed by Jaeger {\it et al} \cite{jae} fails at low vibration intensities
because it does not take into account cooperative phenomena which leads to
formation of avalanches. The avalanche dynamics should be the main
contribution to the relaxation of the pile angle at low vibration
intensities, close to the static limit.

In the present paper the angle relaxation of a sandpile under vibrations is
investigated. Avalanches are modelled by a mean-field approximation
represented in a Bethe lattice. Typical characteristics of complex systems:
long time tails relaxation, aging and fractal distributions of time
constants are obtained by simple numerical simulations. The main goal of
this work if to study the sandpile dynamics far from equilibrium.

In a static sandpile sand grains are in equilibrium unless the angle of the
pile overcomes its critical value determined by the static friction
coefficient. However, the grains in a pile under vibration are vibrating
around certain quasiequilibrium positions but from time to time the
amplitude of its vibrations becomes large. These large amplitude motions
carry as a consequence a convective motion in the bulk and the relaxation of
the pile angle.

At low intensity vibrations large amplitude grain motions are rare events
but from time to time a grain may jump from its quasiequilibrium position.
If a surface grain jumps it will fall through the slope of the pile until it
collides with another sand grains down the slope. After this collision the
initial grain may be trapped with those grains or some of them may fall
through the slope. If the last possibility happens then each grain
''surviving'' the collision will fall through the slope as the initial grain
did. This image of an avalanche as an initial object that consecutively
drags another resembles a branching process for which the Bethe lattice
representation seems to be natural \cite{zap}.

However, representing the avalanche dynamics in a Bethe lattice is a mean
field approach to the problem, it neglects correlations between branches.
Notwithstanding, if the pile angle is below is critical value the avalanches
will be rare events and, in case of occurrence, will be very sparse. Thus,
such an approximation will be acceptable for analyzing the long time
relaxation of the pile angle below its critical value, which is the subject
of this paper. Away from the critical point the mean field approximation
works quite well.

Let us represent the avalanche as a cascade in the Bethe lattice as follows.
Firstly, we start with a single node, which could represent in this case a
grain. In a further step $F$ will emerge with probability $p(\theta )$,
depending on the pile angle $\theta $. This operation of generation of $F$
identical particles starting from one is repeated in the next step to each
node of the new group, and so on. If the percolation process overcomes a
given length (that of the border of the pile) those nodes beyond the limit
constitute the avalanche. If it does not, there would be no avalanche since
the cascade was stopped before reaching the base of the pile (frustrated
avalanche). By avalanche size we take the number of nodes, of the
corresponding Bethe lattice, in the last step.

After an avalanche the sandpile {\em autorganizes itself} with the new number
of grains (i.e. a new slope is calculated with the remaining grains). The
occurrence of avalanches will carry as consequence a decrease in the number
of grains in the pile and, therefore, a decrease of $p(\theta )$. As more
avalanches take place less probable is the occurrence of a new avalanche.
Thus to characterize this feedback mechanism a relation between $p$ and the
number of grains in the pile $N$ is needed.

The drag probability $p(\theta )$ is a function of the slope. Its value is
determined by the competition of two contrary forces: the gravity which
conspires against the stability of the slope and the friction which favours the slope stability. Since the slope forms an angle $\theta $ with the horizontal plane the component of the gravity force in the slope direction will be larger with the increase of this angle, varying from zero to a maximum value when $\theta$ goes from zero to $\pi /2$. Therefore it is plausible to assume that the contribution of the gravity and, therefore, the tendency of falling down the slope is proportional to $\sin \theta $. On the contrary, the resistance to this tendency given by the static friction decreases with decreasing the pile angle, varying from a maximum value to zero when $\theta $ goes from zero to $\pi /2$. Hence, it is also plausible to assume that the resistance to the falling down is proportional to $\cos \theta $.

Thus the slope dependence of $p$ should be given through the ratio of both
tendencies $\sin \theta /\cos \theta =\tan \theta $. Based on this
hypothesis we propose the exponential relation
\begin{equation}
p(\theta )=\exp (-A/\tan \theta )\ ,  \label{eq:1}
\end{equation}
where $A$ is a parameter determined by the gravitational field, the
friction, and vibration intensity. Notices that $p(0)=0$ and $p(\pi /2)=1$.

On the other hand, as it is well known the number of grains in a pile with
slope angle $\theta $ is given by 
\begin{equation}
N=Bx^3\tan \theta  \label{eq:2}
\end{equation}
where $B$ is a geometrical factor and $x=D/d$ is the ration between a
characteristic size of the pile base $D$ and sand grain $d$. Combining
equations \ref{eq:1} and \ref{eq:2} it is obtained

\begin{equation}
p(N)=\exp (-cx^3/N)\ .  \label{eq:3}
\end{equation}
where $c=AB$. Thus equation \ref{eq:1} relates the dragging probability with
the number of grains in the pile.

The parameter $c$ should not depend on the size of the system. It must be a
function of parameters describing the vibrations, amplitude $a$ and
frequency $\omega$, and of the gravitational field acceleration $g$. The
only nondimentional combination of this magnitudes is given by the ratio
between the vibration acceleration $a\omega^{2}$ and $g$ and, therefore, $c=c(a\omega^2/g)$. This conclusion obtained from dimensional analysis is
corroborated by experiments on sandpiles under vibrations \cite{eve} which
shown that the ratio $a\omega^{2}/g$ is the relevant parameter.

For a given value of $c$ and $x$ there is a critical number of grains in the
pile $N_c$. This value can be found recalling that in the Bethe lattice the
critical value of $p$ for percolation exists is $1/F$, resulting 
\begin{equation}
N_c=cx^3/\ln F\ .  \label{eq:4}
\end{equation}
In a static sandpile (i.e. no vibrations) this value corresponds with the
number of grains in the pile at the angle of repose~\cite{zap}.

Another magnitude of interest is the penetration length, the number of steps 
$n(\theta )$ in the Bethe lattice for an avalanche take place. This
magnitude must be proportional to the length of the pile slope and,
therefore, should be given by the expression 
\begin{equation}
n(\theta )=x/\cos \theta  \label{eq:4a}
\end{equation}
Notice that the possible existence of a geometrical factor in equation \ref
{eq:4a} may be absorbed in $x$, redefining $B$ in equation \ref{eq:2}.
Moreover, the relation between $n$ and $N$ may be easily obtained using
equation \ref{eq:2}.

Equations \ref{eq:3} and \ref{eq:4a} relates the parameters of the pile with
those of its Bethe lattice representation. They were obtained here in a
different way than in \cite{zap}. Other dependencies between the dragging
probability and the number of grains in the sandpile may be proposed.
Notwithstanding, as it is discussed latter, the precise form of this
functional dependence is not too relevant.

The numerical experiment of relaxation is performed as follows. We start
with a certain number of $N$. In a first step, we test if an avalanche takes
place using the Bethe lattice representation. If it does then we recalculate
the value of $p(N)$ by simply resting the size of the avalanche to $N$ and
using equation \ref{eq:3}. Then this step is repeated again and again.

If avalanches are considered as instantaneous the number of steps is a
measure of time. This approximation is valid for low vibrations intensities.
In this case grain jumps which trigger avalanches are rare events and,
therefore, the time between to successive grain jumps will be much larger
than the duration of avalanches.

\begin{figure}
\narrowtext
\centerline{\epsfbox{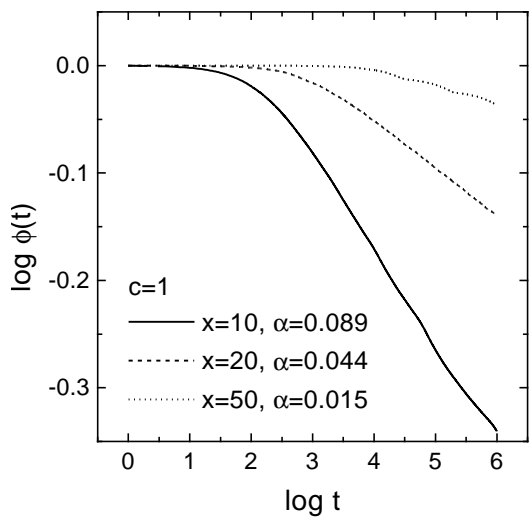}}
\caption{Normalized relaxation $\phi(t)=N(t)/N(t=0)$ of the number of
grains in the pile for different values of $x$. The initial condition is $%
N(t=0)=N_{c}$ in all cases. Insets show the values of $\alpha$ obtained from
the fit to the power tail $t^{-\alpha}$.}
\label{fig:1}
\end{figure}

\begin{figure}
\narrowtext
\centerline{\epsfbox{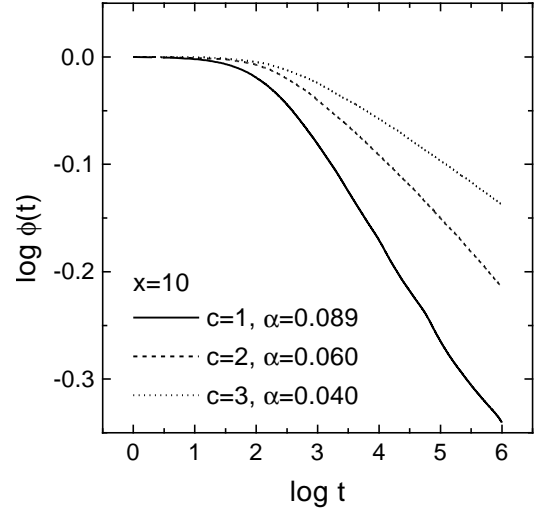}}
\caption{Normalized relaxation $\phi(t)=N(t)/N(t=0)$ of the number of
grains in the pile for different values of $c$. The initial condition is $N(t=0)=N_{c}$ in all cases. Insets show the values of $\alpha$ obtained from the fit to the power tail $t^{-\alpha}$.}
\label{fig:2}
\end{figure}

The normalized decay of $N$ in time for different values of $x$ and $c$ is
shown in figures \ref{fig:1} and \ref{fig:2}. In all cases we took as initialcondition $N(t=0)=N_c$, the number of grains at the angle of repose in the corresponding static limit, and we have averaged over $10000$ realizations. As it can be seen the relaxation follows a fast decay at short times but it is slower than an exponential at long times. The log-log plot reveals a straight line at long times characteristic of a long time tail $t^{-\alpha }$. Insets show the values of $\alpha $ obtained from the fit to this power tail, having relative errors of the order of $0.01$.

The characteristic exponent $\alpha $ decreases towards zero with increasing 
$x$, i.e. the relaxation is slower for larger sandpiles. It is expected that
it is strictly zero for $x\rightarrow \infty $, which means that an infinite
sandpile will be in equilibrium. This limit may be the equivalent of the
ferromagnetic state of magnetic materials where domain are macroscopic
structures having finite magnetization. On the other hand, with increasing $c$, $\alpha $ decreases towards zero and since the relaxation should be
slower for smaller vibrations intensities we may conclude that $c$ must
increase with decreasing the vibration intensity. Hence, $c$ is a
monotically decreasing function of $a\omega ^2/g$.

\begin{figure}
\narrowtext
\centerline{\epsfbox{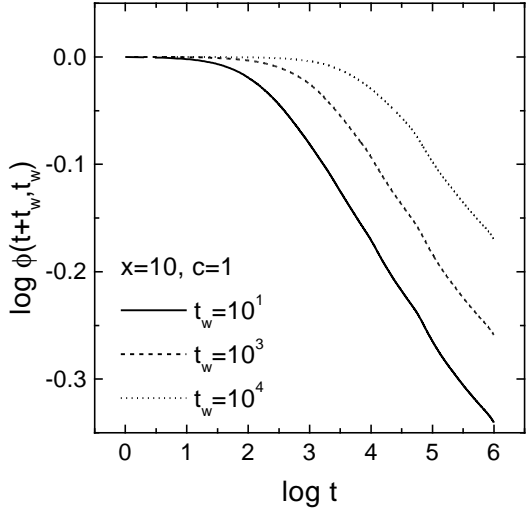}}
\caption{Aging. Normalized relaxation function $\phi(t+t_{w},t_{w})=
N(t+t_{w})/N(t_{w})$ taken at three different initial times $t_{w}$.}
\label{fig:3}
\end{figure}

The existence of aging in our model is illustrated in figure \ref{fig:3}. We have plotted the normalized relaxation at different ages (steps) of the system(simulation) by taking as initial time $t=t_w$. As it can be see the
relaxation becomes slower with increasing the age of the system, in
agreement with experiments in structural and spin glasses \cite{bou}. Thus,
since the angle relaxation becomes slower with the age of the pile then it
will never reach an equilibrium angle and, therefore, properties like
translational invariance and the fluctuation dissipation theorem do not hold 
\cite{bou}.

Associated with these slow relaxation dynamics and aging phenomena we expect
to observe a wide distribution of time between avalanches. As time increases
the time between two consecutive avalanches $\Delta t$ becomes larger and
larger, because after an avalanche the occurrence of a new avalanche becomes
smaller. Therefore, it is expected that the mean time between avalanches
diverges as $t\rightarrow \infty $. Hence, the distribution of time between
avalanches $n(\Delta t)$ should satisfies the asymptotic behavior for large $\Delta t$
\begin{equation}
n(\Delta t)\sim \ \Delta t^{-1-\beta }  \label{eq:5}
\end{equation}
with $0<\beta <1$.

\begin{figure}
\narrowtext
\centerline{\epsfbox{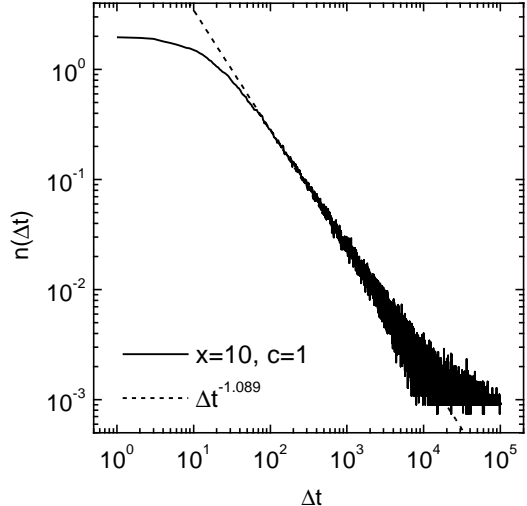}}
\caption{Distribution of time between avalanches. The dashed line is a
power tail with exponent $-1-\alpha$, $\alpha$ being the exponent of the
long time tail shown in figure 1.}
\label{fig:4}
\end{figure}

This hypothesis is confirmed in our simulations. Figure \ref{fig:4} shows the distribution of time between avalanches for $x=10$ and $c=1$. It is
approximately constant for small values and then it decays following a power
law in more than two decades. The plateau at small values of $\Delta t$ is
associated to the rapid decay observed at short times while the tail for
large $\Delta t$ should be related to the long time tail. Thus, there should
be some connection between the distribution of time between avalanches and
the long time relaxation.

In fact, if we assume that the relaxation is given by the superposition of
exponential relaxations, the relaxation time being the time between
avalanches, the temporal decay of $N$ will be given by 
\begin{equation}
N(t)=N_0\int_0^\infty d\Delta t\ n(\Delta t)\ \exp (-t/\Delta t)\ .
\label{eq:6}
\end{equation}
Then, for long times equation \ref{eq:6} gives the asymptotic behavior 
\begin{equation}
N(t)\sim t^{-\beta }\ .  \label{eq:7}
\end{equation}
Thus $\alpha =\beta $, the exponent of the long time tail and the
characteristic exponent of the distribution of time between collisions are
mutually determined. This result is corroborated in figure \ref{fig:4} where it can be see that a power decay $\sim \Delta t^{-1-\alpha}$ fits quite well the distribution of time between avalanches for large $\Delta t$. Long time
tails relaxation and fractal distributions of time constants are the same
phenomena manifested in different windows \cite{va}.

\begin{figure}
\narrowtext
\centerline{\epsfbox{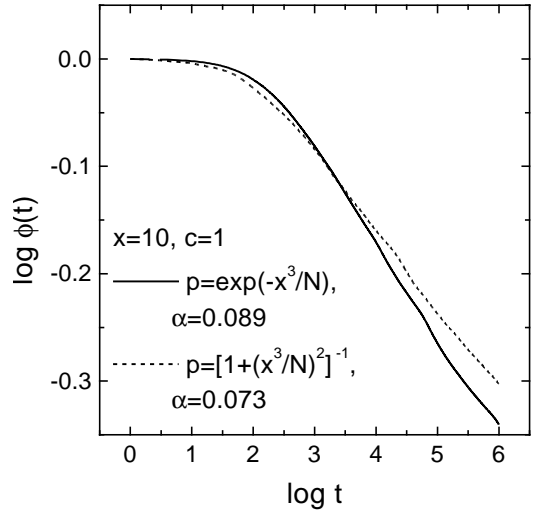}}
\caption{Normalized relaxation $\phi(t)=N(t)/N(t=0)$ of the number of
grains in the pile for two different functional dependences $p=p(N)$. The
initial condition is $N(t=0)=N_{c}$ in both cases. Insets show the values of 
$\alpha$ obtained from the fit to the power tail $t^{-\alpha}$.}
\label{fig:5}
\end{figure}

Finally we want to mention that these simulations where also carried out
assuming other functional dependences between the dragging probability and
the number of grains in the pile $p(N)$. In all cases the results where
qualitatively similar to those presented here using equation \ref{eq:3},
reflecting that the precise dependence is not determinant. For instance, in
figure \ref{fig:5} the normalized relaxation for two different functional forms of $p(N)$ is plotted. In both cases the long time relaxation follows a long time tail but with different exponents. The qualitative form of the relaxation does not depends on the detailed form of $p(N)$, i.e. on the detailed nature of the system. Nevertheless, the form of $p(N)$ does determine the value of the long time tail exponent $\alpha $ suggesting that different universality classes are possible.

We conclude that the introduction of a Bethe lattice representation for the
avalanches and a feedback mechanism describes quite well the principal
features of the relaxation in sandpiles under low intensity vibrations. The
proposed representation leads to long time tails relaxation, aging,and
fractal distributions of time constants, which are characteristic properties
of the dynamics of complex systems out from the equilibrium. The slow
relaxation dynamics and related properties are a consequence of the feedback
mechanism, but the detailed nature of this feedback is not relevant for the
qualitative behavior.

We want to acknowledge Professors Ernesto Atshuler, Roberto Mulet and
Stephan Boettcher for useful comments and helpful discussion.

This work was partially supported by the {\it Alma Mater} prize, given by
The University of Havana, and by the International Centre of Theoretical
Physics (ICTP).

\end{multicols}

\end{document}